\documentclass[12pt]{iopart}
\usepackage{graphicx}
\usepackage{iopams}
\usepackage{amsthm,amssymb}

\newtheorem{theorem}{Theorem}[section]

\newcommand{\bR}{{\mathbb R}}

\theoremstyle{remark}

\eqnobysec

\begin{document}

\title{A method to construct refracting profiles}

\author{N Alamo\dag\ and C Criado\ddag
}

\address{\dag\ Departamento de Algebra, Geometria y Topologia,
Universidad de Malaga, 29071 Malaga, Spain }

\address{\ddag\ Departamento de Fisica Aplicada I, Universidad de
Malaga, 29071 Malaga, Spain }

\date{}

\eads{\mailto{nieves@agt.cie.uma.es}, \mailto {c\_criado@uma.es}}

\begin{abstract}
We propose an original method for determining suitable refracting
profiles between two media to solve two related problems: to
produce a given wave front from a single point source after
refraction at the refracting profile, and to focus a given wave
front in a fixed point. These profiles are obtained as envelopes
of specific families of Cartesian ovals. We study the
singularities of these profiles and give a method to construct
them from the data of the associated caustic.

\end{abstract}

\maketitle

\section{Introduction}
Suppose that we are given a surface $R$ of $\bR^3$, which
separates two media where the light travels with different
velocities $v_1$ and $v_2$, and a point $F$ in the region
corresponding to $v_1$. In what follows, $R$ will be called
refracting profile and $F$ point source of light. Let $W$ be
another surface in the region corresponding to $v_2$, such that
its normal lines coincide with the straight lines that emerge from
$F$ after refraction at $R$. The surfaces $W$ with these
properties are called wave fronts. A classical optical problem is
to determine the wave front from the data of the point source $F$
and the refracting profile $R$. It can be solved by using the
Snell's law, or equivalently, the Fermat's principle.

In this paper we consider the inverse problem, that is, given the
wave front $W$ and the source point $F$, we want to construct a
refracting profile $R$ such that $W$ is the wave front obtained
from a spherical wave front emanating from $F$ after refraction at
$R$. The same profile will solve the problem of focusing the wave
front $W$ in the point $F$.

We will give a method based on the optical properties of the
Cartesian ovals (section 2), to obtain a family $\{ R^a\} $ of
refracting profiles parametrized by a non-negative real number
$a$, with the property of producing normal lines to $W$ after
refraction at each profile of the rays that emerge from $F$. Each
profile $R^a$ will be the envelope of a family of Cartesian ovals
(see figure 1). A similar method have been used in
\cite{antiorthotomics} to construct reflecting profiles as
envelopes of certain families of conics.

Next section contains the details of the construction of the
profiles $R^a$ and the proof that, under very general conditions,
these profiles are immersed surfaces of $\bR^3$ that have
singularities only if they contain centres of curvature of the
wave front $W$. Moreover, we prove that the singularities of the
families of the profiles $R^a$ sweep out the caustic of the wave
front $W$. In section 3 we study the optical physical sense of
these profiles.

Finally, in section 4 we show an interesting relation between the
profiles $R^a$ and the caustic of the given wave front, which also
allows us to construct them from the caustic.

\section{Construction of the refracting profiles}
Let us remember some properties of the Cartesian ovals. The ovals
of Descartes or Cartesian ovals were introduced by Descartes in
1637 in his Dioptrique, dedicated to the study of light
refraction. A Cartesian oval is the locus of the points from which
the distances $r_1$ and $r_2$ to two fixed points $F_1$ and $F_2$,
called foci, verify the bipolar equation
\begin{equation}\label{oval}
a_1 r_1  + a_2 r_2 = k,
\end{equation}
where $a_1$, $a_2$ and $k$ are constants. Observe that this
equation includes the bifocal definition of conics for the
particular cases of $k> 0$ and $a_1=a_2>0$ (ellipse), or $a_1=
-a_2>0$ (hyperbola).

A Cartesian oval has a third focus $F_3$, and the oval can be
defined by any two of the foci (see \cite{Lockwood}). In
particular, when $\vert a_1/a_2\vert$ goes to $1$, the third focus
goes to infinity and we get the conics.

The so-called complete Cartesian oval is the set of curves
associated to the bipolar equation
\begin{equation}\label{completeoval}
a_1 r_1  \pm a_2 r_2 = \pm k \quad  (k>0).
\end{equation}
Only two of the four equations obtained from these double signs
are not empty. If $a_1\not= a_2$, they are closed curves, one
interior to the other (see figure~2 with $F_1 =F$ and $F_2 = x$).
These two curves intersect only when two foci coincide. In this
case we get the so-called Lima\c con of Pascal. See
\cite{Lockwood} for a detailed study of these ovals. Reference
\cite{Rabal} contains some applications of these ovals to
holography.

\begin{figure}
\begin{center}
\includegraphics{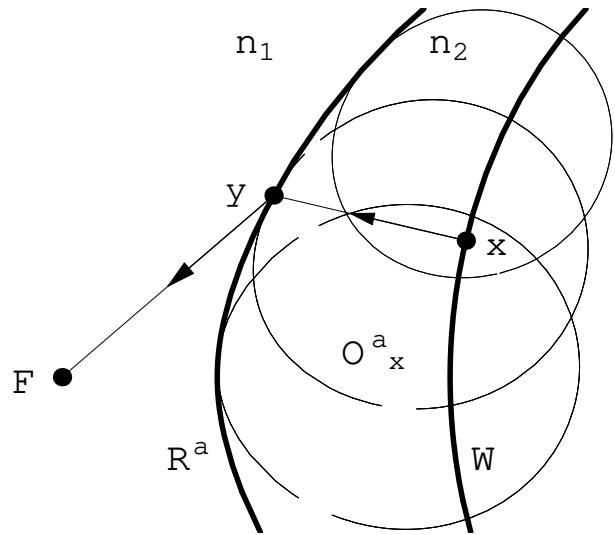}
\caption{The wave front $W$ focuses in $F$ after refraction at the
profile $R^a$. This profile is the envelope of a family of
interior Descartes ovals $O_x^a$ with foci $F$ and $x\in W$; $n_1$
and $n_2$  are the refractive indices.}
\end{center}
\label{Figuno}
\end{figure}

A complete Cartesian oval is symmetric with respect to the
straight line determined by the foci $F_1$ and $F_2$. Then, if we
consider $F_1$ and $F_2$ belonging to $\bR^3$, the bipolar
equation \eref{completeoval} is the equation of a surface of
revolution in $\bR^3$ with axis of revolution $F_1F_2$. Since the
intersection of this surface with any plane containing $F_1$ and
$F_2$ is a complete Cartesian oval, we have called this surface
complete Cartesian oval of revolution.

The fact that makes interesting Cartesian ovals of revolution in
the context of geometrical optics is the following. Consider a
surface $R$ separating two media with refractive indices $n_1$ and
$n_2$, and two points $F_1$ and $F_2$ in medium  $n_1$ and  $n_2$
respectively. If we want that a ray coming from  $F_1$ goes to
$F_2$, then the point $P$ where the ray cross $S$ must verify that
$n_1 \vert F_1P\vert +  n_2 \vert F_2P\vert $ be constant, but
this shows that $S$ is part of a Cartesian oval of revolution, as
we see looking at Eqn.\eref{completeoval}. Because the refractive
indices $n_1$ and $n_2$ are related with the light velocities
$v_1$ and $v_2$ by $n_i=c/v_i$,   $i=1,2$, where $c$ is the light
vacuum velocity, the above condition is equivalent to say, that
all the rays spend the same time going from $F_1$ to $F_2$, and
this is the Fermat principle. We will see in Section 3, that only
some relative position of the foci with respect to the complete
oval of revolution and some portions of this oval have physical
sense.

 In what follows let  $O_x^a$ denote the complete Cartesian oval
 of revolution with foci $F$ and $x\in W$, satisfying
 Eqn.\eref{completeoval} for the parameters $a_1$ and $a_2$ equal
  to the refractive indices $n_1$ and $n_2$ respectively, and
 the optical path length between $F$ and
$x$, $k$, equals to $2a$. Let
  $R^a$ be the envelope of the family  $\{ O_x^a\}_{x\in W}$.
We will give an explicit parametrization of $R^a$ and from it we
will determine which portion have physical sense as refracting
profile.

From now on we assume that $F$ is the origin of $\bR^3$, so any
point $y\in O_x^a$ must verify:
\begin{equation}\label{concreoval}
n_1 \vert y\vert \pm  n_2 \vert y -x \vert = \pm 2a.
\end{equation}

Therefore, for any $a\geq 0$, we can define the family of complete
Cartesian ovals of revolution $\{ O_x^a\}_{x\in W}$ by mean of the
maps $f_a : W\times \bR^3 \rightarrow \bR$ given by
$
f_a(x,y) = n_1 \sqrt {y \cdot y} \pm n_2  \sqrt {(y - x)\cdot (y -
x)} \pm 2a$, \, so that for any fixed $x\in W$  the zero set of
the map $f_{a,x}: \bR^3 \rightarrow \bR$ given by $f_{a,x} (y) =
f_a(x,y)$, is the Cartesian oval $O_x^a$.

According to the definition of the envelope of a two-parameter
family of surfaces (see  Refs. \cite{Courant,Goursat}), the
envelope $R^a$ of the family $\{ O_x^a\}_{x\in W}$ is the set of
points $y \in  \bR^3$ such that exists $t=(t_1,t_2)\in U$
verifying the equations:
\begin{equation}\label{eqenvelope}
 f_a(x(t_1,t_2),y)=0, {{\partial f_a}\over{\partial
t_1}}(x(t_1,t_2),y) = 0, {{\partial f_a}\over{\partial
t_2}}(x(t_1,t_2),y) = 0
\end{equation}
for $t=(t_1,t_2)$ being the standard coordinates in $\bR^2$ and
$x:U\subset \bR^2 \rightarrow \bR^3$ a parametrization of W.

The envelope $R^a$ is therefore obtained by eliminating $t_1$ and
$t_2$ between the following equations
\begin{eqnarray}
f_a(x(t),y) & = & 0\\ \label{eqpartial} {{\partial
f_a}\over{\partial t_i}}(x(t),y) & = & \pm n_2 {{y-x(t)} \over
{\vert y - x(t) \vert}} \cdot \big(-{{\partial x}\over{\partial
t_i}}\big) = 0, \quad i= 1,2.
\end{eqnarray}

Then, equation \eref{eqpartial} becomes
\begin{eqnarray}
\label{eqnormalline} (y-x(t))\cdot {{\partial x}\over{\partial
t_i}} = 0, \quad  i= 1,2,
\end{eqnarray}
which, is also the equation of the line $r_n(x)$, normal to $W$ at
$x(t)$, and thus it follows that
\begin{equation}\label{eqintersection}
R^a= \bigcup_{x\in W}O_x^a\cap r_n(x).
\end{equation}
This proves that the normal line to $W$ at $x$ intersects the
envelope $R^a$ at the points $y(x)$ where the envelope itself is
tangent to the Cartesian oval $O_x^a$ (see figure~2).

In order to give an explicit parametrization of $R^a$, we suppose
that $W$ is oriented and we take $n(x)$ to be the outward unitary
normal vector field to $W$. According to (\ref{eqintersection}),
$R^a$ can be given by the parametric equation
\begin{equation}\label{eqnormal}
y= x + \lambda (x) n(x)
\end{equation}
where $\lambda (x)$ has to be determined by the condition that
$y\in O_x^a$, that is, $y$ has to verify \eref{concreoval}.

A straightforward calculus gives four $\lambda$'s:
\begin{equation}\label{eqlambdas}
\lambda_{i,j} = {{2 a n_2 - (-1)^i n_1^2 (x\cdot n) + (-1)^{i+j}
\sqrt{\Delta_i} }\over {n_2^2 - n_1^2}}, \quad  i= 1,2,
\end{equation}
which are defined only when $\Delta_i = (2 a n_2 - (-1)^i n_1^2
(x\cdot n))^2 - (n_2^2 - n_1^2)(4 a^2 - n_1^2 x^2)$ is positive.
Then $R^a$ can be decomposed in four sheets $R^a = \cup_{i, j =
1,2} R_{i,j}^a$, where $R_{i,j}^a$ is parametrized by
\begin{equation}\label{eqnormalij}
y= x + \lambda_{i,j} (x) n(x)
\end{equation}

Figure~2 illustrates these four sheets in a 2-dimensional example.
 \begin{figure}
\begin{center}
\includegraphics{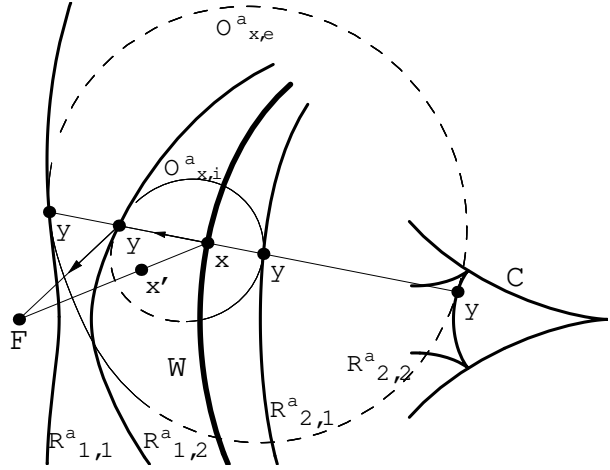}
\caption{The wave front $W$ and the four profiles $R_{i,j}^a$
($i,j=1,2$), which are the envelopes of the interior ($O_{x,i}^a$)
and exterior ($O_{x,e}^a$) Descartes ovals. These ovals have foci
$F$ and $x\in W$; $x'$ is the third focus. The singularities of
the profile $R_{2,2}^a$ are on the caustic $\mathcal{C}$ of the
wave front.}
\end{center}
\label{Figdos}
\end{figure}

Let us now see that the map $y_{ij}:U\subset \bR^2 \rightarrow
\bR^3$ given by $y_{ij}(t)= x(t) + \lambda_{i,j} (x(t)) n(x(t))$,
($i,j=1,2$) is an immersion, so that $R^a = \cup_{i, j = 1,2}
R_{i,j}^a$ is an immersed surface of $\bR^3$.

To simplify the notation we avoid the indices $i,j$ in
\eref{eqnormalij}. To see that $y$ is an immersion we have to
verify that the vectors
\begin{equation}
{{\partial y}\over{\partial t_k}} = {{\partial x}\over{\partial
t_k}} + \lambda {{\partial n}\over{\partial t_k}} + {{\partial
\lambda}\over{\partial t_k}}n,\quad k= 1,2,
\end{equation}
are linearly independent.

If we chose a suitable parametrization in $W$, we can assume that
the vectors ${{\partial x}\over{\partial t_1}}, {{\partial
x}\over{\partial t_2}}$ are orthonormals.  With respect to the
orthonormal base $\{ {{\partial x}\over{\partial t_1}}, {{\partial
x}\over{\partial t_2}}, n\}$ of $\bR^3$, the two first components
of the vectors ${{\partial y}\over{\partial t_1}}, {{\partial
y}\over{\partial t_2}}$ form a $(2\times 2)$- matrix given by:
\begin{equation}\label{eqmatrix}
\left( {{\partial x}\over{\partial t_i}}\cdot {{\partial
y}\over{\partial t_j}} \right) = \left( \delta_{ij} + \lambda
{{\partial x}\over{\partial t_i}}\cdot {{\partial n}\over{\partial
t_j}} \right) = \left( \delta_{ij} - \lambda {{\partial^2
x}\over{\partial t_i \partial t_j}} \cdot n \right) ,
\end{equation}
where for the second identity we have used that \/  $ 0 =
{{\partial}\over{\partial t_j}}\left({{\partial x}\over{\partial
t_i}}\cdot n \right) = {{\partial x}\over{\partial t_i}}\cdot
{{\partial n}\over{\partial t_j}} + {{\partial^2 x}\over{\partial
t_i \partial t_j}} \cdot n$. This matrix is singular if and only
if $1/\lambda$ is an eigenvalue of the matrix $\left( {{\partial^2
x}\over{\partial t_i
\partial t_j}} \cdot n \right)$
which is the matrix associated to the second fundamental form of
$W$ at $x(t)$ with respect to the base  $\{ {{\partial
x}\over{\partial t_1}}, {{\partial x}\over{\partial t_2}}\}$ of
the tangent plane to $W$ (see p. 141 of \cite{DC}). Therefore, its
eigenvalues are the principal curvatures of $W$ at $x(t)$.

It follows that the matrix (\ref{eqmatrix}) is singular if and
only if $1/\lambda$ is a principal curvature of $W$ at $x(t)$, or
equivalently, $y(t)$ is a centre of principal curvature of $W$ at
$x(t)$. Then we have proved the following theorem:

\begin{theorem}\label{1} Suppose that $R^a$ does not contain any
centre of principal curvature of $W$. Then
 $R^a$ is an immersed surface of
$\bR^3$.
\end{theorem}

According to the above theorem, the singularities of the
refracting profiles $R^a$ are the points $y\in R^a$ which are
centres of principal curvature  of $W$. The locus $\mathcal{C}$,
of the centres of principal curvature of the wave front $W$, or
equivalently, the envelope of the normal lines to $W$, is called
the caustic of $W$.

In what follows we only consider wave fronts whose caustics are
non-degenerate in the sense that each point $c$ of the caustic
corresponds to an unique point $x$ of the wave front.

It is known that when the wave front propagates its singularities
slide along the caustic, see \cite{Arnold1,Arnold2}. It was proved
in \cite{antiorthotomics} that a similar result holds regarding
the singularities of the reflecting profiles. In the next theorem
we prove that an analogous result is valid for the profiles $R^a$.

\begin{theorem}\label{singularities}
For each $c\in {{\mathcal{C}}}$ of $W$ there are two refracting
profiles $R^{a_1}$ and $R^{a_2}$ such that $c$ is a singularity of
both of them.
\end{theorem}
We express this fact saying that the singularities of the family
$\{R^a\}_{a\ge 0}$ sweep out the caustic ${\mathcal{C}}$ of $W$
twice.
\begin{proof} For each point $c\in {{\mathcal{C}}}$ take $a_1=\frac{\vert
n_1 \vert c\vert \ +  n_2 \vert c - x\vert \vert}{2} $ and
$a_2=\frac{\vert n_1 \vert c\vert \ - n_2 \vert c - x\vert
\vert}{2} $.
\end{proof}

Figure~3 gives an illustrative example of how the singularities of
the profiles $R^a$ sweep out the caustic ${\mathcal{C}}$.

\begin{figure}
\begin{center}
\includegraphics{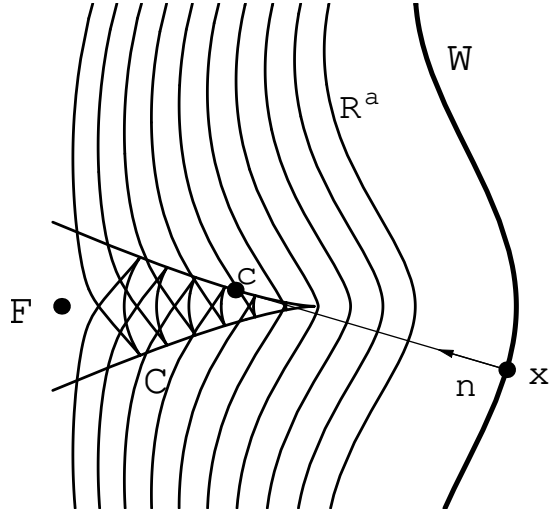}
\caption{The singularities of the profiles $R^a$ sweep out the
caustic $\mathcal{C}$ of the wave front $W$.}
\end{center}
\label{Figtres}
\end{figure}

\section{Physical sense of the different profiles}

We will study now which sheets of $R^a$ are adequate as refracting
and focusing profiles. That is, which sheets produce the wave
front $W$ after refracting at them a spherical wave front
emanating from $F$, or equivalently, focus $W$ in $F$. First of
all, we will give an explicit parametrization of the family of
complete Cartesian ovals $\{ O_x^a\}_{x\in W}$.

To reduce the problem to the 2-dimensional case, we consider  for
every $x\in W$ the plane $\pi (F, x)$ determined by $F$, $x$, and
$n(x)$. Now if we take $F$ as pole and $Fx$ as polar axis, the
polar equation $r = r(x,\varphi)$ for a point $y\in O_x^a$ is
determined by the condition:
\begin{equation}\label{polareq}
\vert n_1 r \pm n_2 r_2\vert = 2a
\end{equation}
where $r = \vert y \vert$, $r_2 = \vert y - x\vert$, and $n_1\not=
n_2$.

On the other hand, $r$ and $r_2$ are, together with $\vert x\vert$
the sides of a triangle. Then:
\begin{equation}\label{triangle}
r_2^2 = r^2 + \vert x \vert^2 - 2 r \vert x\vert \cos \varphi.
\end{equation}

Using \eref{polareq} and \eref{triangle} we get two solutions:
\begin{equation}\label{twosheets}
r_{\pm}(x,\varphi) = {{2 a n_1 - n_2^2 \vert x\vert \cos \varphi
\pm \sqrt{\Delta (x,\varphi)}} \over {n_1^2 - n_2^2}}
\end{equation}
where
\begin{equation}\label{Delta}
\Delta(x,\varphi) = (2 a n_1 - n_2^2 \vert x\vert \cos \varphi)^2
- (n_1^2 - n_2^2) (4 a^2 - n_2^2 \vert x \vert^2).
\end{equation}

Consider now a cartesian coordinate system of the plane $\pi
(F,x)$, $F,X,Y$, with origin at F. With respect to this coordinate
system the cartesian equations associated to $r_{\pm}(x,\varphi)$
are:

\begin{equation}\label{yi}
y_{\pm} = (r_{\pm}(x,\varphi)\cos(\varphi+\theta),
r_{\pm}(x,\varphi)\sin(\varphi+\theta)),
\end{equation}
where $\theta$ is the angle determined by $x$ and $FX$.

In figure~4(a) we have represented an example with these two
solutions.
 It illustrates also the following discussion.

 The solution $r_+$
(dashed line in figure~4(a)), has two parts: one in the interior
oval
 $O_{x,i}^a$, and the other in the exterior oval
 $O_{x,e}^a$. They correspond to the values of $\varphi$ for
which $\Delta (x,\varphi)>0$, and are separated by other two
regions where $\Delta (x,\varphi)<0$. So there are four values of
$\varphi$ for which $\Delta (x,\varphi)=0$, as can be deduced from
\eref{Delta}.

These four values of $\varphi$ are the corresponding to the values
of the angle between $FX$ and the four tangent lines to the
interior $O_{x,i}^a$ and the exterior $O_{x,e}^a$ ovals passing
through $F$.

For the solution $r_-$ (solid line in figure~4(a)), we have an
analogous decomposition.

Then both the interior and exterior ovals are made up of two
pieces, one corresponding to $r_+$ (dashed line in figure~4(a))
and the other corresponding to $r_-$ (solid line in figure~4 (a)).

When $F$ is in the interior of the ovals, $\Delta_{}$ never
vanishes, so in this case, the exterior and interior ovals
coincide with the ovals $r_+$ (dashed line in figure~4(b)) and
$r_-$ (solid line in figure~4(b)) respectively.

A sheet of $R^a$ could have physical sense as refracting or
focusing profile only when $F$ and $W$ are at different sides of
the tangent plane to the sheet at any point of it, and a point $y$
being in that sheet must verify $n_1 \vert y\vert + n_2 \vert y -
x\vert = 2 a$. On the other hand, for any $x\in W$, a point $y$ in
the interior oval $O_{x,i}^a$ verifies (see \cite{Lockwood})
\begin{equation}\label{interioroval}
n_1 \vert y\vert + n_2 \vert y - x\vert = 2 a
\end{equation}
and a point  $y$ of the exterior oval $O_{x,e}^a$ verifies
\begin{equation}\label{exterioroval}
- n_1 \vert y\vert + n_2 \vert y - x\vert = 2 a.
\end{equation}

From condition \eref{interioroval} we know that the refracted ray
at the point $y\in O_{x,i}^a$ of a ray emanating from $F$,
 passes through
$x$ (see figure~2). We know also by (2.7) that the normal line to
$W$ at $x$ intersects the envelope $R_{1,2}^a$ at the points
$y(x)$ where the envelope itself is tangent to the oval $O_x^a$.
This fact together with the above observation
 means that the rays produced from the point source $F$ by
refraction at $R_{1,2}^a$ are the normal rays to $W$. Therefore we
have proved the following result, which can be considered the main
result of this paper.

\begin{theorem}\label{3} Let $W$, $F$, and $R_{1,2}^a$ be as defined above.
Then the profile $R_{1,2}^a$ refracts the rays emerging from $F$
into the normal rays to $W$.
\end{theorem}

In other words the theorem says that the wave front $W$ can be
obtained from a spherical wave front emerging from $F$ after
refraction at $R_{1,2}^a$, or equivalently the profile $R_{1,2}^a$
focuses the wave front $W$ in $F$.

We can prove with an analogous reasoning that the sheet
$R_{2,2}^a$ has a physical sense when we consider $F$ as a virtual
source point. Let us assume that the profile $R_{2,2}^a$ separates
two media with refractive indices $n_2$ in the side containing $F$
and $n_1$ in the other side, and that both $F$ and $W$ are at the
same side of the tangent plane to the sheet for any point of it
(see figure~4(a)). Then we have the following:

\begin{theorem}\label{4} The wave front $W$, when propagates
towards  $R_{2,2}^a$ and after refraction at it, produces a
spherical divergent wave front ${\mathcal{S}}$ with centre $F$.
\end{theorem}
\noindent{\it Remark.} If we reverse the direction of propagation
of the wave front, the above theorem can equivalently be stated
as: A convergent to $F$ spherical wave front ${\mathcal{S}}$, when
propagates toward $R_{2,2}^a$, and after refraction at it,
produces the wave front $W$. We say in this case that $F$ is a
virtual source of $W$ for the refracting profile $R_{2,2}^a$.
\begin{proof} Consider a point $z\in {\mathcal{S}}$, so it
verifies $\vert z\vert = R$ (constant). From the refraction law
the optical path length $n_2 \vert y - x\vert + n_1 \vert z -
y\vert$ from $x\in W$ to $z\in {\mathcal{S}}$ must be constant.
Moreover, since $\vert z - y\vert = R - \vert y\vert $ then $-n_1
\vert y \vert + n_2 \vert y - x\vert$ must be constant, and this
is just the equation that satisfy the points $y$ in the exterior
ovals $\{O_{x,e}^a\}_{x\in W}$, whose envelope is $R_{2,2}^a$.
\end{proof}

Using together theorems \ref{3} and \ref{4} we can derive an
interesting optical property. Consider a wave front $W$, a source
point $F$, and a parameter $a$ such that $W$ and $F$ are at
different sides of $R_{1,2}^a$. Suppose given refractive indices
$n_1 < n_2$ such that the region delimited by $R_{1,2}^a$ and
$R_{2,2}^a$ is filled with a medium of refractive index $n_2$, and
the exterior of this region with refractive index $n_1$. Then a
spherical divergent wave front emerging from $F$ produces, after
refraction at $R_{1,2}^a$  the wave front $W$, and after another
refraction at $R_{2,2}^a$, produces a propagation of the initial
spherical wave front (see figure~4(a)). This process is a special
kind of the known as do-nothing machines (\cite{DONO}), and it can
have useful applications to optic, acoustic and radar devises.

The physical sense of the different sheets of the profiles $R^a$
depends on the kind of ovals (exterior or interior), whose
envelope is the sheet, as well as the relative position of the
considered sheet with respect to the wave front $W$ and the source
point $F$. As has been shown before, when $W$ and $F$ are at
different sides with respect to the profile $R_{1,2}^a$, then
$R_{1,2}^a$ and $R_{2,2}^a$ have got the suitable properties of
refracting profiles.

\begin{figure}
\begin{center}
\includegraphics{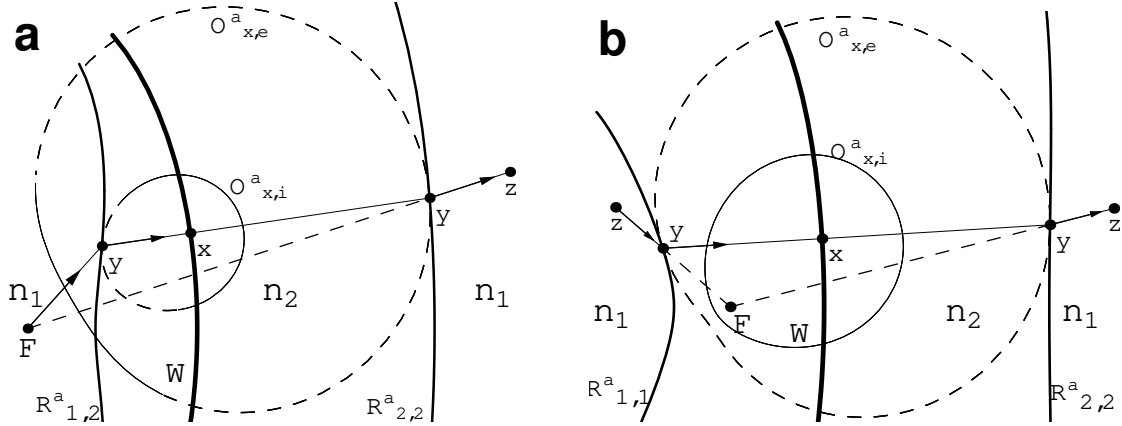}
\caption{{\bf{(a)}} If $F$ is in the exterior of the ovals, a
spherical wave front emerging from $F$ produces, after refraction
at $R_{1,2}^a$, the wave front $W$, and, after a posterior
refraction at $R_{2,2}^a$, appear again the initial spherical wave
front. {\bf{(b)}} If $F$ is in the interior of the ovals, a
spherical wave front converging to $F$ produces, after refraction
at $R_{1,1}^a$, the wave front $W$, and a posterior refraction at
$R_{2,2}^a$ produces a spherical divergent wave front with centre
$F$. }
\end{center}
\label{Figcuatro}
\end{figure}

There is another profile sheet with physical sense, namely the
profile $R_{1,1}^a$. Consider a value of the parameter $a$ such
that $F$ and $W$ are in the same region with respect to the
tangent plane to this sheet, i.e. $n(x) \cdot y > 0$, for $n(x)$
pointing to the medium $n_1$ (see figure~4(b)). Reasoning as
before, we obtain that under these condition:

\begin{theorem}\label{5} Any spherical wave front ${\mathcal{S}}$,
convergent to $F$ from the medium of $n_1$, after refraction at
$R_{1,1}^a$ produces the wave front $W$.
\end{theorem}

Equivalently, the theorem says that the wave front $W$ after
refraction at $R_{1,1}^a$, produces a spherical divergent wave
front ${\mathcal{S}}$ with centre $F$.

A nice optical property can be deduced from theorems  \ref{4} and
\ref{5}.

Consider that the region delimited by $R_{1,1}^a$ and $R_{2,2}^a$
is filled with a medium of refractive index $n_2$, and the
exterior of this region with refractive index $n_1$. Then a
spherical wave front converging to $F$ and after refraction at
$R_{1,1}^a$, produces the wave front $W$, and subsequent
refraction at $R_{2,2}^a$ gives a divergent spherical wave front
with centre $F$ (see figure~4(b)). This property can also have
useful applications.

Notice that if we want to have physical sense, in all the above
discussion we must exclude the region of the profile in which
there are total reflection, that is, we have to exclude the points
$y$ of the profile such that the corresponding incidence angle
$\theta_1$ does not give a real value for the angle of refraction
$\theta_2$. It occurs when light is propagated from an optically
denser medium into one which is optically less dense, i.e. when
$n_2< n_1$, provided that the incidence angle $\theta_1$ exceeds
the critical value $\theta_c$ given by $\sin \theta_c = n_2/n_1$.

\section{Construction of the $R^a$-profiles from the caustic}

In this section we develop a method to construct the
$R^a$-profiles of a given wave front $W$ with respect to a point
$F$, starting from the data of its caustic ${\mathcal{C}}$. As we
set in the above section, we only consider wave fronts whose
caustics are non degenerated in the sense that each point $c$ of a
caustic corresponds to a unique point $x$ of the wave front.

For each regular point $c\in {\mathcal{C}}$ corresponding to a
point $x\in W$ and a  principal curvature radius $\rho$, such that
$c$ can be expressed as $c= x+\rho n(x)$,
 set $a'= a + {\rho \over 2}n_2 $\, and \,
 $a''= \vert a - {\rho \over 2}n_2\vert$.

Consider the families of ovals of revolution $\{ O_c^{a'}\}_{c\in
\mathcal{C}}$ and $\{ O_c^{a''}\}_{c\in \mathcal{C}}$, with foci
$F$ and $c$ and parameters $a'$ and $a''$ respectively.

Then we have:

\begin{theorem}\label{caustics} With the notation above, the
$R^a$-profiles are in the envelopes of the families of complete
ovals of revolution $\{O_c^{a'} \cup O_c^{a''}\}_{c\in
{\mathcal{C}}}$.
\end{theorem}
\begin{proof}
We shall prove the result for the convex regions of the wave
front. The proof for the concave regions is analogous.

\begin{figure}
\begin{center}
\includegraphics{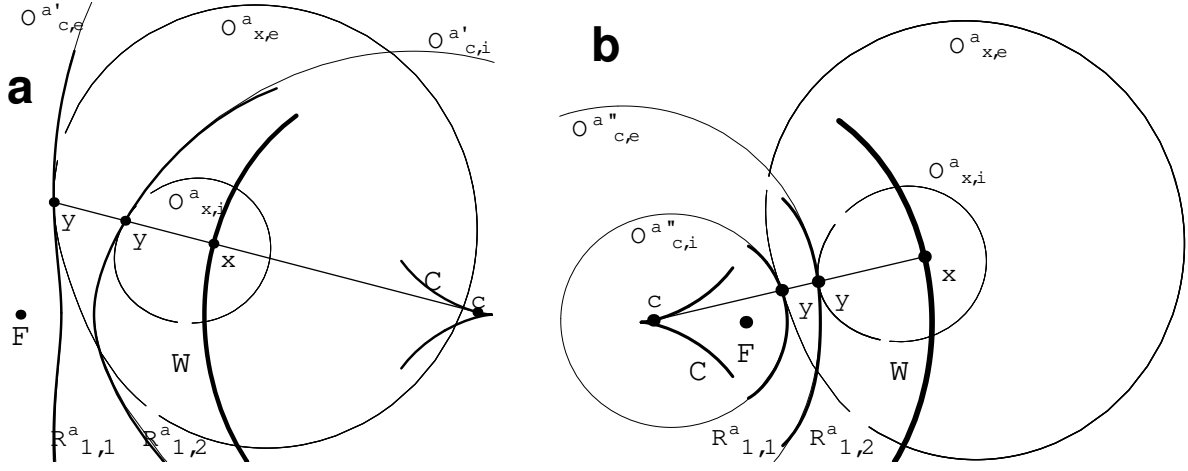}
\caption{{\bf{(a)}} For the case of $W$ convex, $R_{1,1}^a$ and
$R_{1,2}^a$ are the envelopes of the ovals $O_{c,e}^{a'}$ and
$O_{c,i}^{a'}$ respectively, which have foci $F$ and $c\in
\mathcal{C}$. {\bf{(b)}} For a concave $W$, $R_{1,1}^a$ and
$R_{1,2}^a$ are the envelopes of the ovals $O_{c,i}^{a''}$ and
$O_{c,e}^{a''}$ respectively, which have foci $F$ and $c\in
\mathcal{C}$. }
\end{center}
\label{Figcinco}
\end{figure}

In the convex regions the sheets $R_{1,1}^a$, $R_{1,2}^a$,
$R_{2,1}^a$, and $R_{2,2}^a$ are the envelopes of the families of
ovals $\{ O_{c,e}^{a'}\}_{c\in \mathcal{C}}$, $\{
O_{c,i}^{a'}\}_{c\in \mathcal{C}}$, $\{ O_{c,e}^{a''}\}_{c\in
\mathcal{C}}$, and $\{ O_{c,i}^{a''}\}_{c\in \mathcal{C}}$
respectively; here the letters $e$ and $i$ denote exterior and
interior sheet respectively of the corresponding complete oval. We
will centre our attention to the case of refracting profiles, that
is, $R_{1,2}^a$. The proof is similar for the other cases.
Figure~5 illustrates with an example the families of ovals whose
envelopes are $R_{1,1}^a$ and $R_{1,2}^a$. We have to prove that
$R_{1,2}^a$ is the envelope of $\{ O_{c,i}^{a'}\}_{c\in
\mathcal{C}}$. We already know that the refracting profile is the
envelope of the interior ovals $\{ O_{x, i}^a\}_{c\in W}$,
verifying
\begin{equation}\label{eqinterior}
n_1 \vert y \vert + n_2 \vert y - x\vert = 2 a.
\end{equation}
On the other hand the points $y \in  O_{c,i}^{a'}$ verify
\begin{equation}\label{eqovalint}
n_1 \vert y \vert + n_2 \vert y - c\vert = 2 a' = 2 a + \rho n_2 =
2 a + \vert x - c\vert n_2,
\end{equation}
where $c$ is the centre of curvature of $W$ at $x$. We shall show
that $O_{c,i}^{a'}$ and $O_{x,i}^a$ are tangent at the point $y$
in which  $O_{x,i}^a$ is tangent to the refracting profile (see
figure~5(a)).

In order to demonstrate this result it suffices to show that the
intersection between $O_{c,i}^{a'}$ and $O_{x,i}^a$ is only one
point, precisely the point $y$ where $O_{x,i}^a$ intersects the
normal line to $W$ at $x$. This follows from the fact that $y$
verifying both (\ref{eqinterior}) and (\ref{eqovalint}) implies
$\vert y - x\vert = \vert y - c\vert - \vert x - c\vert$, what it
is possible only if $y$ is in the straight line determined $x$ and
$c$. Observe that $\vert y - c\vert > \vert x - c\vert$ since for
a convex wave front the caustic ${\mathcal{C}}$ and $F$ are in
opposite sides with respect to $W$ (see figure~5(a)).

For the concave case, $R_{1,1}^a$, $R_{1,2}^a$, $R_{2,1}^a$, and
$R_{2,2}^a$ are the envelopes of the families of ovals $\{
O_{c,e}^{a''}\}_{c\in \mathcal{C}}$, $\{ O_{c,i}^{a''}\}_{c\in
\mathcal{C}}$, $\{ O_{c,i}^{a'}\}_{c\in \mathcal{C}}$, and $\{
O_{c,e}^{a'}\}_{c\in \mathcal{C}}$ respectively. Figure~5(b)
illustrates this case for $R_{1,1}^a$ and $R_{1,2}^a$.
\end{proof}

\section{Conclusions}
Using envelopes of families of Cartesian ovals of revolution, we
have given an original procedure to solve the following inverse
problem: find a refracting profile $R$ separating two media such
that for a given wave front $W$ and a fixed point source $F$, the
wave front produced by $F$ after refracting at $R$ is precisely
$W$. In fact, with the given procedure we have constructed a
one-parameter family of profiles, each member of this family
having four sheets. We have studied the optical physical sense of
these sheets.

We have proved that these refracting profiles are immersed
surfaces in $\bR^3$, with singularities in the points
corresponding to the centres of curvature of the wave front.
Finally, we have given another way to reconstruct the refracting
profile from the caustic.

The proposed method can be of interest in the study and design of
refracting profiles. The fact that each one of these profiles is
constructed as the envelope of Cartesian ovals of revolution could
be used to design it as a tesselation of smaller ovals profiles.
This composed profile could be a good approximation of the
original one, and therefore could be useful in the construction of
optical, radar, or acoustical devices. We have also found that
with a suitable choice of two of these profiles, it is possible to
construct devises with do-nothing machines properties.

For $n_1=n_2$ the Cartesian ovals are ellipses or hyperbolas, and
some of the sheets of the $R^a$-profiles can be physically
interpreted as suitable reflecting and focusing profiles. From
this point of view, this work can be considered as a
generalization of \cite{antiorthotomics}.

Finally we note that the construction can be easily generalized to
$\bR^n$, $n >3$.

\ack Partial financial support from Spanish grants BFM2001-1825
(N. Alamo) and FQM-192 (C. Criado) is acknowledged.

\end{document}